\documentclass{PoS}
\usepackage{xspace}
\usepackage{amsmath,amssymb,slashed} 
\usepackage{textcomp}  
\providecommand\romanup[1]{\text{#1}}  

\newcommand\raa{R_\text{AA}}
\newcommand\dd{\mathop{}\!\romanup{d}}

\title{Heavy-flavor dynamics in event-by-event viscous hydrodynamic backgrounds}

\ShortTitle{Open heavy flavors with the DAB-mod simulation.}

\author{\speaker{Roland Katz}\\ 
       Instituto de F\'{i}sica, Universidade de S\~{a}o Paulo, C.P. 66318, 05315-970 S\~{a}o Paulo, SP, Brazil\\
        E-mail: \email{rkatz@if.usp.br}}

\author{Caio A.~G.~Prado\\
        Institute of Particle Physics, Central China Normal University, Wuhan, Hubei 430079, China\\
        E-mail: \email{cagprado@mail.ccnu.edu.cn}}

\author{Jacquelyn Noronha-Hostler\\
        Department of Physics and Astronomy, Rutgers University, Piscataway, NJ 08854, USA\\
        E-mail: \email{jacquelyn.noronhahostler@rutgers.edu}}

\author{Alexandre A.~P.~Suaide\\
        Instituto de F\'{i}sica, Universidade de S\~{a}o Paulo, C.P. 66318, 05315-970 S\~{a}o Paulo, SP, Brazil\\
        E-mail: \email{suaide@if.usp.br}}

\author{Jorge Noronha\\
        Instituto de F\'{i}sica, Universidade de S\~{a}o Paulo, C.P. 66318, 05315-970 S\~{a}o Paulo, SP, Brazil\\
        E-mail: \email{noronha@if.usp.br}}

\author{Marcelo G.~Munhoz\\
        Instituto de F\'{i}sica, Universidade de S\~{a}o Paulo, C.P. 66318, 05315-970 S\~{a}o Paulo, SP, Brazil\\
        E-mail: \email{munhoz@if.usp.br}}

\abstract{We investigate the effects of (2+1)d event-by-event fluctuating hydrodynamic backgrounds on the nuclear modification factor and momentum anisotropies of heavy-flavor mesons. Using the state-of-the-art D and B mesons modular simulation code (the so-called \texttt{DAB-mod}), updated recently with heavy-light quark coalescence, we perform a systematic comparison of different transport equations, including two energy loss models and a relativistic Langevin model with two drag parametrizations. We present the resulting D$^0$ meson $\raa$, $v_2$ and $v_3$, using the multiparticle cumulant method, in Pb-Pb collisions at $\sqrt{s_{\rm NN}}=5.02$ TeV and compare them to the latest experimental data. We investigate the $v_2\{4\}/v_2\{2\}$ ratio as a function of centrality for different initial conditions (MCKLN vs. Trento) and different system geometries and sizes (coming from Pb-Pb collisions at $\sqrt{s_{\rm NN}}=5.02$ TeV, spherical and prolate Xe-Xe collisions at $\sqrt{s_{\rm NN}}=5.44$ TeV).}

\FullConference{International Conference on Hard and Electromagnetic Probes of High-Energy Nuclear Collisions\\
		30 September - 5 October 2018\\
		Aix-Les-Bains, Savoie, France}

\begin{document}


\section{Introduction}

The properties of the medium produced in heavy-ion collisions can be studied in a "tomographic" manner using high $p_T$ particles or heavy-flavor hadrons. Due to the large separation of energy scales $E\gg\Lambda_{\rm QCD}$ and $E\gg T$, these hard probes are only produced via pQCD processes at the very beginning of the collision and they do not flow with their hydrodynamical background but propagate through other processes sensitive to the medium properties. Observables in the heavy-flavor sector such as the nuclear modification factor $\raa$ and azimuthal anisotropy coefficients $v_n$ of D and B mesons have been investigated both experimentally and theoretically \cite{Andronic:2015wma}, and there is a certain difficulty to describe both $\raa$ and $v_2$ simultaneously. 
Here we present the 2D Monte Carlo D-And-B MODular simulation code \texttt{DAB-mod}, designed to study the production of open heavy mesons \cite{Prado:2016szr}. Using \texttt{DAB-mod} we study heavy-flavor multiparticle cumulants of azimuthal anisotropies \cite{Betz:2016ayq} and compare results from different transport models within the same background. We investigate the consequence of initial conditions and nuclei geometries on the heavy-quark dynamics and explore new observables which could help to discriminate between the models.


\section{The DAB-mod simulation}

The heavy quarks are evolved on the top of event-by-event relativistic viscous hydrodynamical backgrounds given by v-USPhydro \cite{Noronha-Hostler:2013gga} with viscosity set to $\eta/s = 0.05$. The initial conditions are varied between MCKLN \cite{Drescher:2007ax} - an implementation of the Color Glass Condensate  - and Trento \cite{Moreland:2014oya} - based on entropy deposition mechanism via a "reduced-thickness" function. Both initial conditions models can reproduce experimental data reasonably well in the soft sector \cite{Betz:2016ayq,Alba:2017hhe,Giacalone:2017dud}. 
\vspace{2.5mm}

The initial heavy quarks are spatially distributed following the initial bulk energy densities and in transverse momentum space with the FONLL spectra and random directions. Shadowing and cold nuclear matter effects are not taken into account. For the propagation of the heavy quarks in the medium we use either a parametric energy loss model \cite{Betz:2016ayq} with two different parametrizations or Brownian motion dynamics via the relativistic Langevin equation with two parametrizations of the diffusion coefficients. The energy loss model is defined by $\dd E/\dd x = -f(T,v_\text{Q}) \Gamma_\text{flow}$, where $T$ is the local temperature, $\Gamma_\text{flow}=\gamma[1 - v_\text{flow}\cos(\varphi_\text{Q} -\varphi_\text{flow})]$ the flow factor accounting for the necessary Lorentz boost between the involved frames, with $\gamma = 1/\sqrt{1-v_\text{flow}^2}$, $\varphi_\text{Q}$ the azimuthal direction of the heavy quark and $\varphi_\text{flow}$ the local azimuthal angle of the flow. The two considered parametrizations, $f(T,v_\text{Q}) = \alpha$ \cite{Das:2015ana} and $f(T,v_\text{Q})=\xi T^2$ where $\alpha$ and $\xi$ are free parameters, have been chosen because they could roughly reproduce the high $p_T$ $\raa$ data trend in $\sqrt{s_\text{NN}}=2.76$ TeV Pb-Pb central collisions. For the relativistic Langevin dynamics, we assume the diffusion coefficients to be isotropic and the momentum space diffusion coefficient $\kappa$ to be independent of the heavy-quark momentum $\vec{p}$, such that  $d\vec{p} = -\Gamma(\vec{p})\vec{p} dt + \sqrt{dt} \sqrt{\kappa}\vec{\rho}$, with $\Gamma$ being the drag coefficient and $\vec{\rho}$ the fluctuating force described classically by a white noise. To be able to reach the correct thermal equilibrium, the associated Einstein relation between the diffusion coefficients is given by $\kappa = 2 E\Gamma T= 2T^2/D$, where $D$ is the spatial diffusion coefficient. We perform the necessary Lorentz boosts between the lab and medium cell rest frames at each time step. We consider two parametrizations of the diffusion coefficients: "M\&T" derived within pQCD plus Hard Thermal Loops (HTL) \cite{Moore:2004tg}, given by $D=k_{\rm M\&T}/(2\pi T)$ where $k_{\rm M\&T}$ is a free parameter, and "G\&A" given by $\Gamma=k_{\rm G\&A}\, A(T,p)$, where the drag $A(T,p)$ is derived from a QCD+HTL model with running coupling and optimized gluon propagator \cite{Gossiaux:2008jv} and where $k_{\rm G\&A}$ is a free parameter. We assume that each heavy quark propagates until it reaches a medium cell where $T<T_d$, where $T_d$ is the decoupling temperature, which is varied between 120-160 MeV in order to assess part of the hadronization uncertainties. The free parameters are fixed to central $\raa$ data at $p_T>10$ GeV for $T_d=120$ and $T_d=160$ MeV (we use the same values for Pb-Pb and Xe-Xe collisions here). \vspace{2.5mm}

The hadronization of the heavy quarks is performed either though fragmentation only (Peterson) \cite{Prado:2016szr}, or through heavy-light quark coalescence. The latter is based on an instantaneous projection of the parton states onto hadron states \cite{Dover,Cao:2015hia}, and the derived coalescence probabilities depend on the heavy-quark momentum, the local flow, and the angle between them. To better fit the recently observed heavy-hadron ratios \cite{Acharya:2018hre}, we included two new elements to these probabilities: 1) a thermal factor of the form "$\exp[-(m_\text{excited}-m_\text{ground})/T_d]$" for the ratios to not only be based on spins and colors but also on the hadron masses, leading to a more relevant statistical hierarchy between the different energy states of a hadron type, 2) a baryon factor ($\sim 3$) to compensate in the model certain missing baryon coalescence mechanisms (e.g. coalescence of a meson resonance with a light quark) leading to an enhancement of the baryon/meson ratios. 


\section{The effects of the transport model, decoupling temperature, and coalescence}

\begin{figure}[h!]
  \centering
  \includegraphics[width=0.45\textwidth]{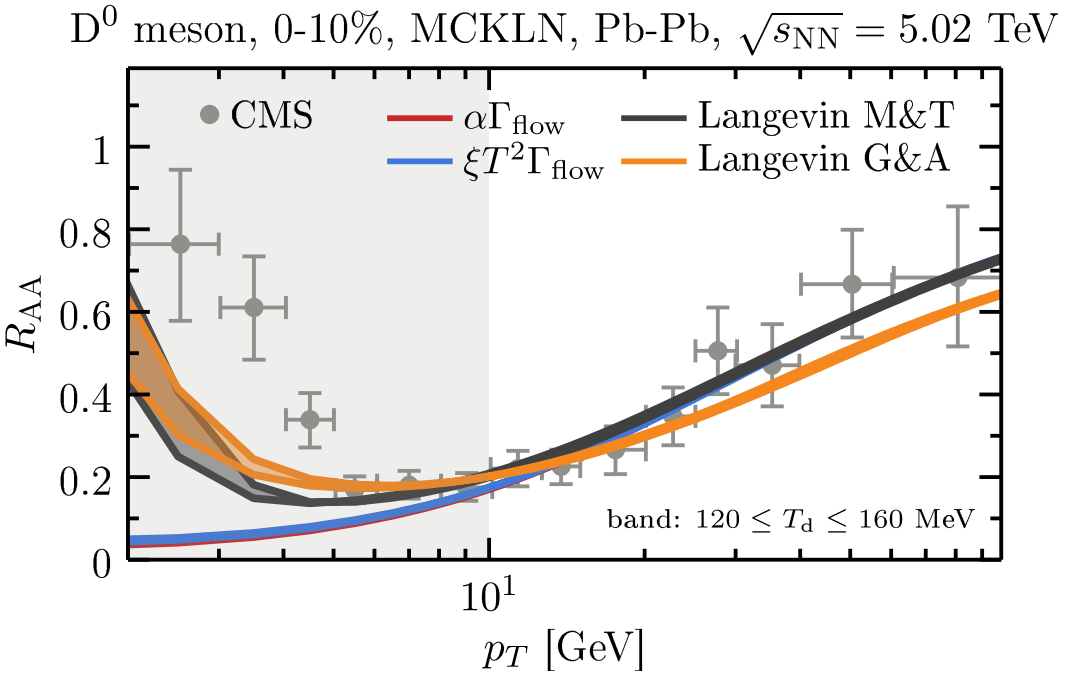}
\hspace{5mm}
  \includegraphics[width=0.465\textwidth]{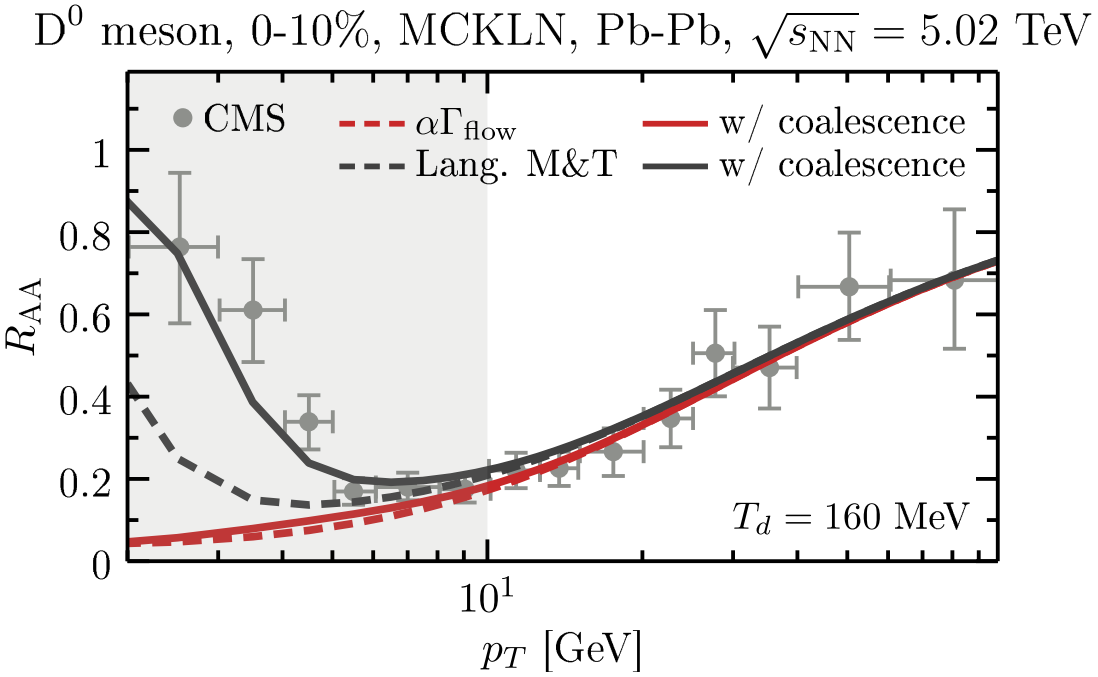}
  \caption{D$^0$ meson $\raa$ for 0-10\% Pb-Pb $\sqrt{s_{\rm NN}}=5.02$ TeV collisions compared to CMS ($|y|<1$) data \cite{Sirunyan:2017xss}. Left: fragmentation only. Right: effect of coalescence. The gray area indicates the $p_T$ region where coalescence and other effects from initial and final stages may be important.}
  \label{fig:RAA_models}
\end{figure}

The $\raa$ is compared in Fig.\ \ref{fig:RAA_models} where the energy loss models fit well at high $p_T$ but clearly miss the data at low $p_T$ whereas the Langevin models can qualitatively capture $\raa$ across all $p_T$ due to the energy gain brought by the fluctuating force. However, only with the inclusion of coalescence in the M\&T Langevin model can we quantitatively reproduce $\raa$ across all $p_T$.

\begin{figure}[h!]
  \centering
  \includegraphics[width=0.45\textwidth]{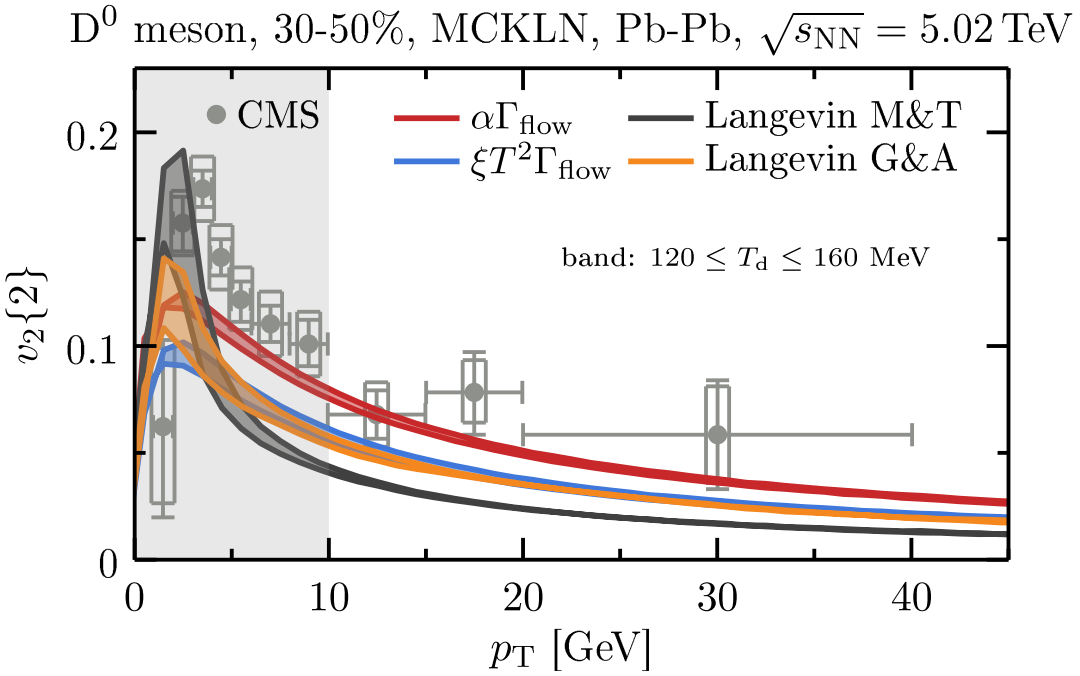}
\hspace{5mm}
  \includegraphics[width=0.45\textwidth]{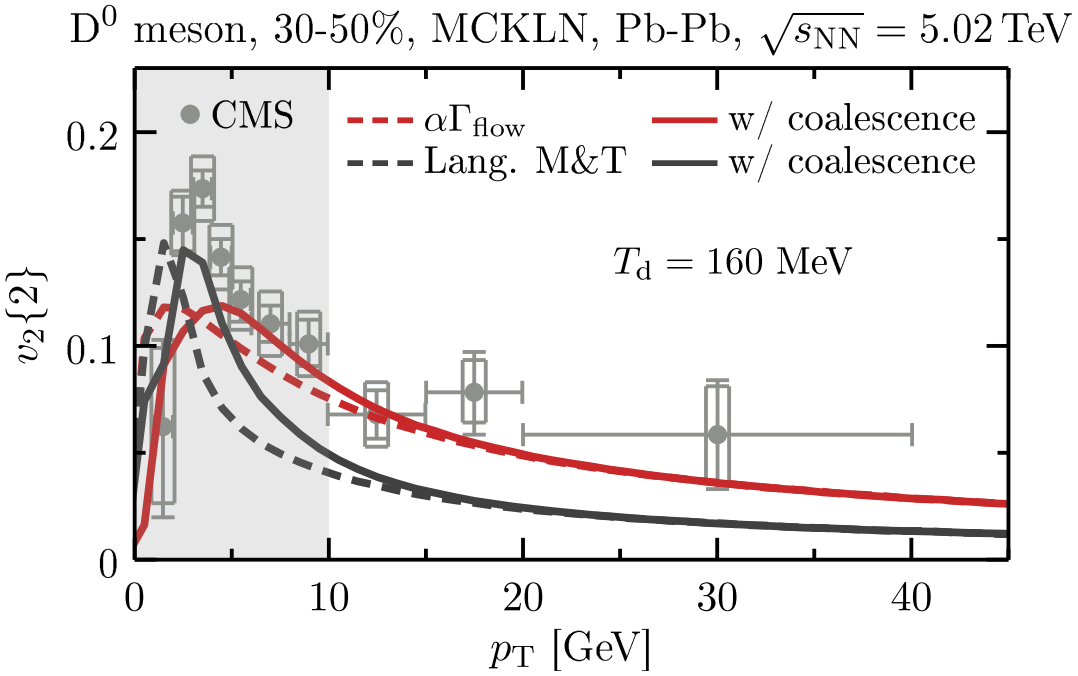}
  \caption{D$^0$ meson elliptic flow coefficient $v_2$ for 30-50\% Pb-Pb $\sqrt{s_{\rm NN}}=5.02$ TeV collisions compared to CMS ($|y|<1$) data \cite{Sirunyan:2017plt}. Left: fragmentation only. Right: effect of coalescence.}
  \label{fig:v2_models}
\end{figure}

\begin{figure}[h!]
  \centering
  \includegraphics[width=0.475\textwidth]{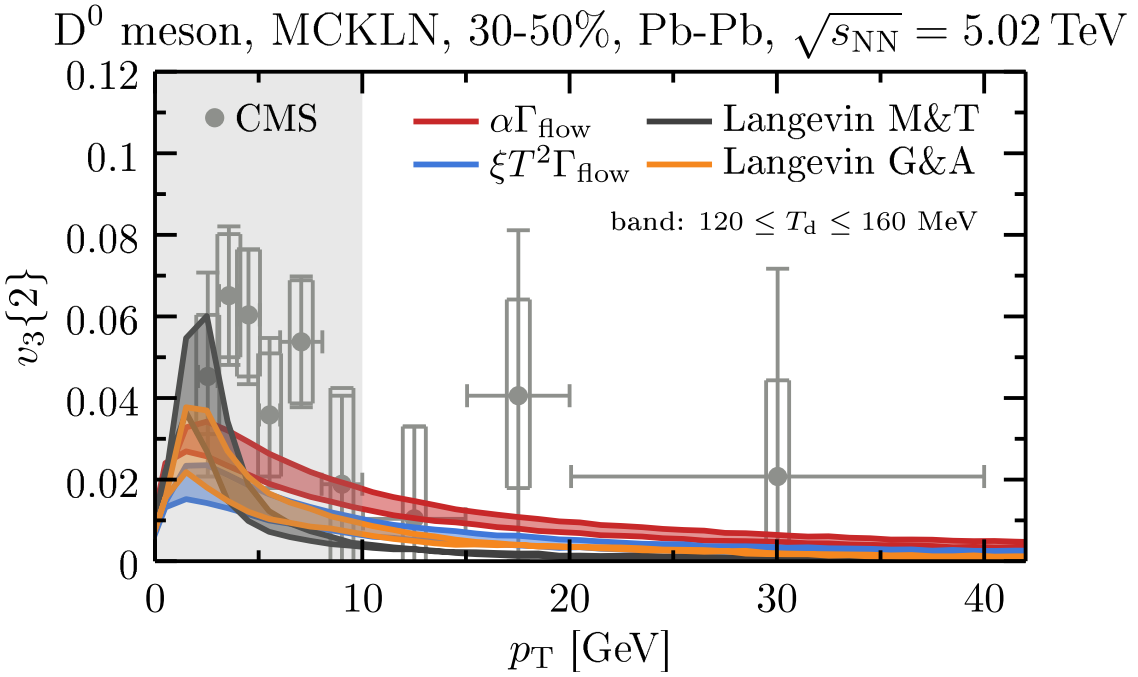}
\hspace{2mm}
  \includegraphics[width=0.472\textwidth]{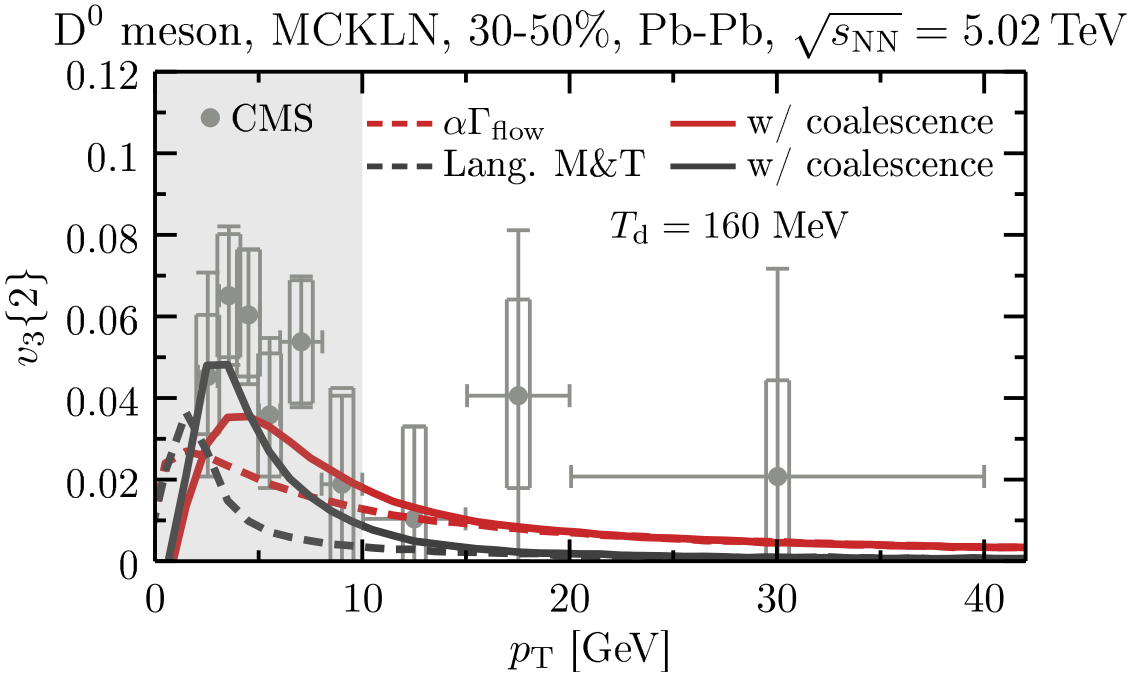}
  \caption{Same as Fig.\ \ref{fig:v2_models} but for the triangular flow coefficient $v_3$.}
  \label{fig:v3_models}
\end{figure}

For the $v_n$ in Fig.\ \ref{fig:v2_models} and \ref{fig:v3_models} we do not find a single model that can quantitatively capture all the experimental data, however, once again the M\&T Langevin model with coalescence has the best fit but it generally underpredicts $v_n(p_T)$ at high $p_T$. In contrast, energy loss models perform best at intermediate to high $p_T$ but underpredict the low $p_T$ behavior. In both cases coalescence is necessary below $p_T<10$ GeV, which shifts the peak in $v_n(p_T)$ to higher $p_T$. This occurs because fragmentation produces a hadron with less momentum than the heavy quark while in coalescence there is a $p_T$ gain from the light quark "thermal" $p_T$ and mass. A lower decoupling temperature $T_d$ leads to a better agreement, which shows that a longer coupling between the heavy quark and the hot medium may compensate missing effects in our model such as the final hadronic rescattering.

\section{The effects of initial conditions and nuclei geometry}

\begin{figure}[h!]
  \centering
  \includegraphics[width=0.43\textwidth]{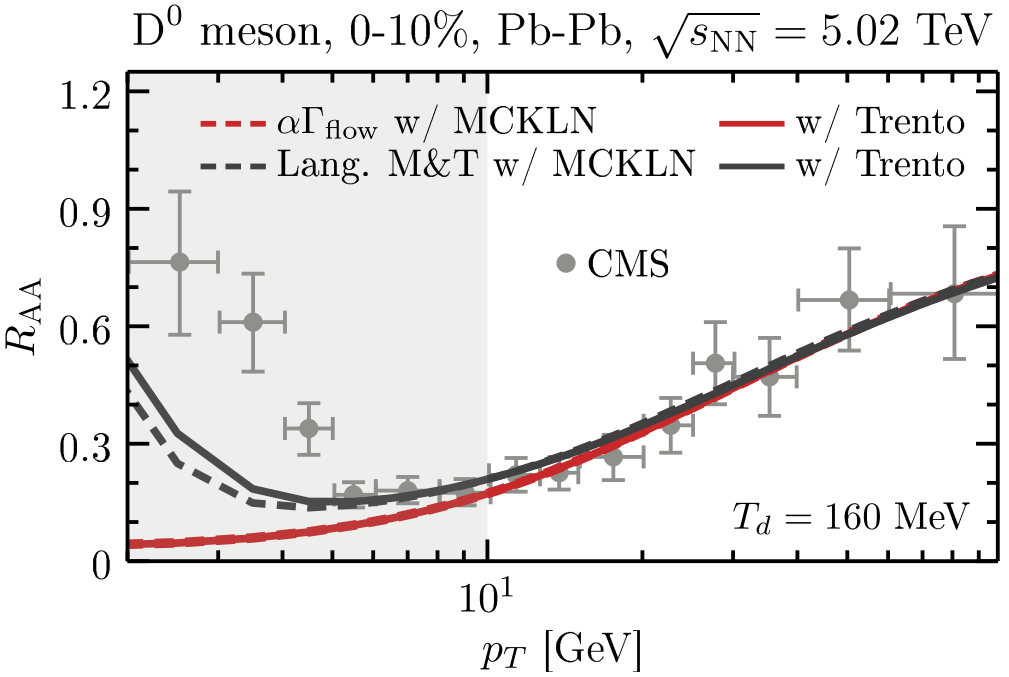}
\hspace{7mm}
  \includegraphics[width=0.432\textwidth]{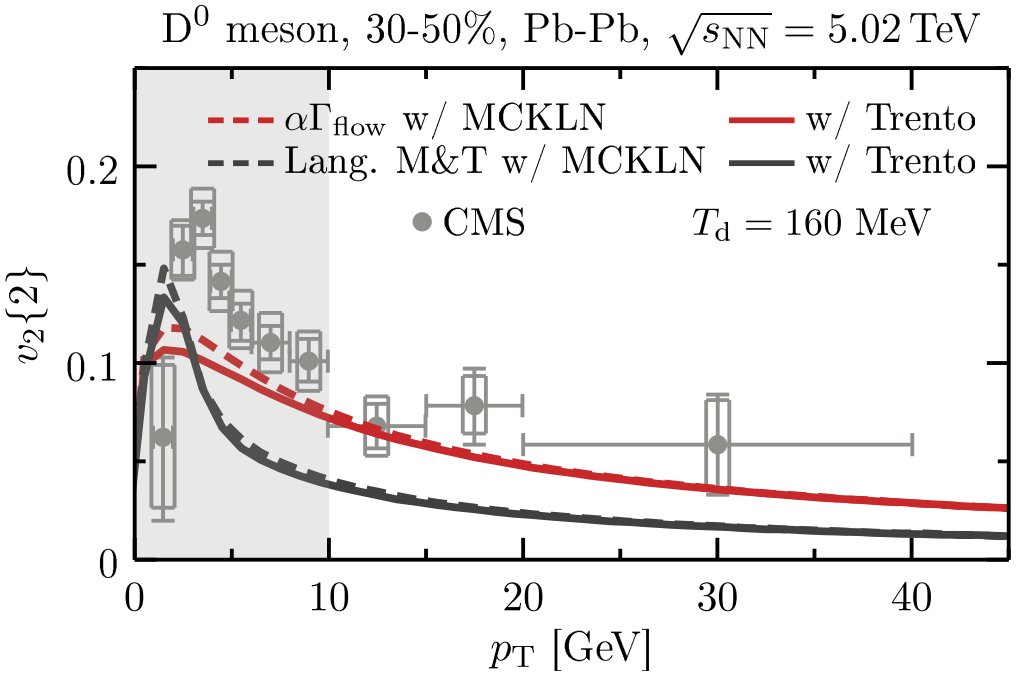}
  \caption{D$^0$ meson $\raa$ (left) and $v_2$ (right) respectively for 0-10\% and 30-50\% Pb-Pb $\sqrt{s_{\rm NN}}=5.02$ TeV collisions compared to CMS ($|y|<1$) data \cite{Sirunyan:2017xss, Sirunyan:2017plt}. Fragmentation only.\vspace{1.5mm}}
  \label{fig:MCKLNvsTrento_models}
\end{figure}

\begin{figure}[h!]
  \centering
  \includegraphics[width=0.43\textwidth]{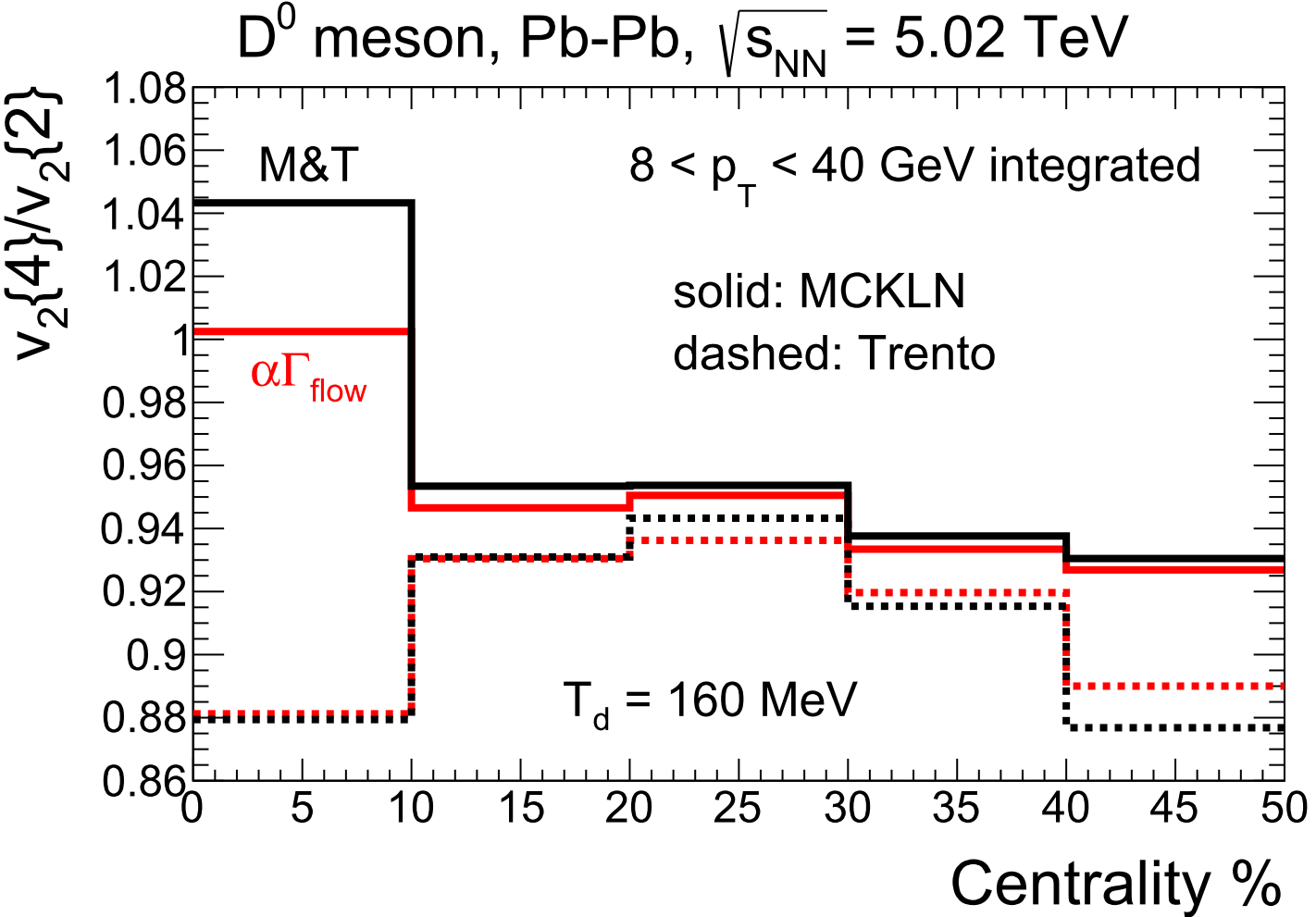}
\hspace{8mm}
  \includegraphics[width=0.428\textwidth]{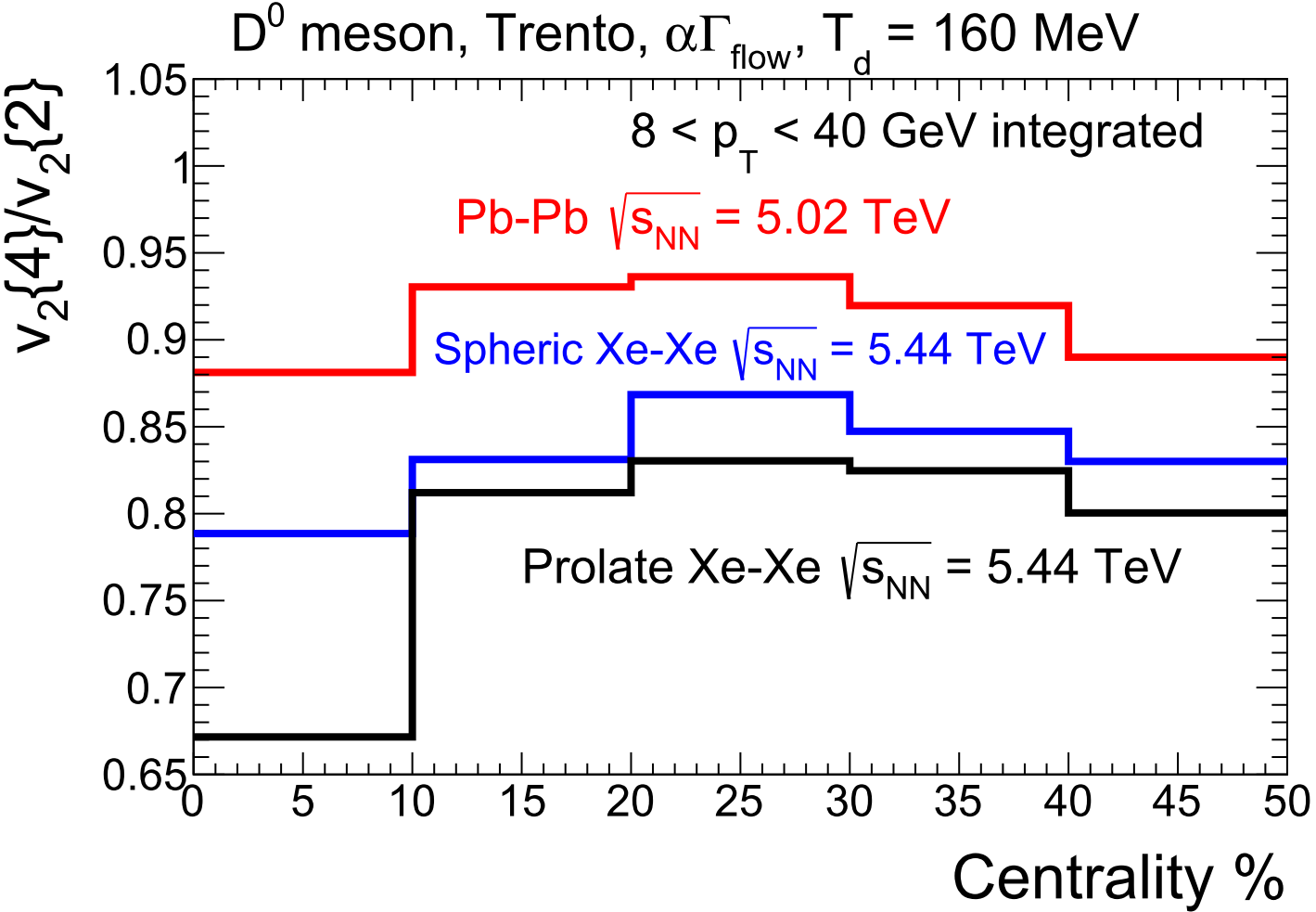}
  \caption{D$^0$ meson $v_2\{4\}/v_2\{2\}$ ratio integrated over $8 < p_T < 40$ GeV. Dependence on the type of initial fluctuations (left) and deformation of Xenon (right). Fragmentation only.}
  \label{fig:Cumulant_ratio}
\end{figure}

Because we fix our free parameters at $p_T>10$ GeV, the choice in initial conditions does not play a large role for $\raa$ and $v_2$ in Fig.\ \ref{fig:MCKLNvsTrento_models}. Generally, Trento produces a slightly smaller $v_2$ at low $p_T$.  Nevertheless, as shown in Fig.\ \ref{fig:Cumulant_ratio} (left), it has an important impact on the trend of the D$^0$ meson $v_2\{4\}/v_2\{2\}$ ratio dependence on centrality, whereas the latter is observed to be mostly independent of the heavy-quark transport model, mass, momentum, decoupling temperature and collision energy $\sqrt{s_{\rm NN}}$ (not shown here). The trend of the Trento curve is very similar to the one observed in the soft sector \cite{Giacalone:2017dud}. Additionally, as shown in Fig.\ \ref{fig:Cumulant_ratio} (right), the "overall" value of this ratio depends strongly on the geometry of the colliding nuclei (size and deformation), whereas the $v_2$ itself not much (not shown here).   

\section{Conclusions}

{\raggedright Comparing the different models, we observed the "multi-scale" behavior of heavy-flavor physics: Langevin models are better suited at low $p_T$ whereas the simple constant energy loss is more relevant in the high $p_T$ regime. Coalescence is observed to improve our results by shifting the low $p_T$ peaks. The $v_2\{4\}/v_2\{2\}$ ratio is mainly dependent on the type of initial fluctuations for its centrality dependence and shows a strong sensitivity to the size and deformation in the nuclei. \vspace{2.5mm}}\\
{\raggedright \emph{Acknowledgements}: The authors thank FAPESP (under grants 2016/17435-8 and 2017/05685-2), CNPq and NSFC (under grant No. 11521064) for financial support. J.N.H. acknowledges the support of the Alfred P. Sloan Foundation and the Office of Advanced Research Computing (OARC) at Rutgers, The State University of New Jersey for providing access to the Amarel cluster and associated research computing resources that have contributed to the results reported here.}



\begin{thebibliography}{99}

\bibitem{Andronic:2015wma} 
  A.~Andronic {\it et al.},
  Eur.\ Phys.\ J.\ C {\bf 76}, no. 3, 107 (2016)
\vspace{-2.3mm}

\bibitem{Prado:2016szr} 
  C.~A.~G.~Prado, {\it et al.},
  Phys.\ Rev.\ C {\bf 96}, no. 6, 064903 (2017)
\vspace{-2.3mm}


\bibitem{Betz:2016ayq} 
  B.~Betz {\it et al.},
  Phys.\ Rev.\ C {\bf 95}, no. 4, 044901 (2017)
\vspace{-2.3mm}

\bibitem{Noronha-Hostler:2013gga} 
  J.~Noronha-Hostler {\it et al.},
  Phys.\ Rev.\ C {\bf 88}, no. 4, 044916 (2013)
\vspace{-2.3mm}

\bibitem{Drescher:2007ax} 
  H.~J.~Drescher and Y.~Nara,
  Phys.\ Rev.\ C {\bf 76}, 041903 (2007)
\vspace{-2.3mm}

\bibitem{Moreland:2014oya} 
  J.~S.~Moreland, J.~E.~Bernhard and S.~A.~Bass,
  Phys.\ Rev.\ C {\bf 92}, no. 1, 011901 (2015)
\vspace{-2.3mm}

\bibitem{Alba:2017hhe} 
  P.~Alba, et al,
  Phys.\ Rev.\ C {\bf 98}, no. 3, 034909 (2018)
  \vspace{-2.3mm}

\bibitem{Giacalone:2017dud} 
  G.~Giacalone {\it et al.}
  Phys.\ Rev.\ C {\bf 97}, no. 3, 034904 (2018)
\vspace{-2.3mm}


\bibitem{Das:2015ana} 
  S.~K.~Das, F.~Scardina, S.~Plumari and V.~Greco,
  Phys.\ Lett.\ B {\bf 747}, 260 (2015)
\vspace{-2.3mm}

\bibitem{Moore:2004tg} 
  G.~D.~Moore and D.~Teaney,
  Phys.\ Rev.\ C {\bf 71}, 064904 (2005)
\vspace{-2.3mm}

\bibitem{Gossiaux:2008jv} 
  P.~B.~Gossiaux and J.~Aichelin,
  Phys.\ Rev.\ C {\bf 78}, 014904 (2008)
\vspace{-2.3mm}

\bibitem{Dover} C. B. Dover, U. Heinz, E. Schnedermann, and J. Zimanyi, Phys. Rev. C {\bf 44}, 1636 (1991).
\vspace{-2.3mm}

\bibitem{Cao:2015hia} 
  S.~Cao, G.~Y.~Qin and S.~A.~Bass,
  Phys.\ Rev.\ C {\bf 92}, no. 2, 024907 (2015)
\vspace{-2.3mm}

\bibitem{Acharya:2018hre} 
  S.~Acharya {\it et al.} [ALICE Collaboration],
  JHEP {\bf 1810}, 174 (2018)
\vspace{-2.3mm}

\bibitem{Sirunyan:2017xss} 
  A.~M.~Sirunyan {\it et al.} [CMS Collaboration],
  Phys.\ Lett.\ B {\bf 782}, 474 (2018)
\vspace{-2.3mm}

\bibitem{Sirunyan:2017plt} 
  A.~M.~Sirunyan {\it et al.} [CMS Collaboration],
  Phys.\ Rev.\ Lett.\  {\bf 120}, no. 20, 202301 (2018)



\end{thebibliography}
\end{document}